# Orbital-Order Driven Ferroelectricity and Dipolar Relaxation Dynamics in Multiferroic GaMo$_4$S$_8$

K. Geirhos[1], S. Krohns[1], H. Nakamura[2], T. Waki[2], Y. Tabata[2], I. Kézsmárki[1], and P. Lunkenheimer[1,*]

[1]Experimental Physics V, Center for Electronic Correlations and Magnetism, University of Augsburg, 86159 Augsburg, Germany
[2]Department of Materials Science and Engineering, Kyoto University, Kyoto 606-8501, Japan

We present the results of broadband dielectric spectroscopy of GaMo$_4$S$_8$, a lacunar spinel system that recently was shown to exhibit non-canonical, orbitally-driven ferroelectricity. Our study reveals complex relaxation dynamics of this multiferroic material, both above and below its Jahn-Teller transition at $T_{JT}$ = 47 K. Above $T_{JT}$, two types of coupled dipolar-orbital dynamics seem to compete: relaxations within cluster-like regions with short-range polar order like in relaxor ferroelectrics and critical fluctuations of only weakly interacting dipoles, the latter resembling the typical dynamics of order-disorder type ferroelectrics. Below the Jahn-Teller transition, the onset of orbital order drives the system into long-range ferroelectric order and dipolar dynamics within the ferroelectric domains is observed. The coupled dipolar and orbital relaxation behavior of GaMo$_4$S$_8$ above the Jahn-Teller transition markedly differs from that of the skyrmion host GaV$_4$S$_8$, which seems to be linked to differences in the structural distortions of the two systems on the unit-cell level.

## I. INTRODUCTION

In recent years, several new and intriguing microscopic mechanisms generating polar order were discovered, leading to such remarkable phenomena as electronic ferroelectricity [1,2,3], multiferroicity [4,5], or topologically protected domain patterns [6,7]. A prominent example is the ferroelectric order that was found to arise below the Jahn-Teller (JT) transitions of several materials [8,9,10,11,12,13]. At a JT transition, long-range orbital order is established due to the lifting of the electronic degeneracy by lattice distortions. While orbital and polar order usually are decoupled, obviously in some cases the orbital ordering can trigger ferroelectric polarization, which is believed to be of mixed electronic and ionic nature [8,9,10,11,12,13]. Very recently, lacunar spinels were revealed to be an important material class that is prone to such orbitally driven polar order [12,13,14,15,16]. Moreover, several of these materials were found to be multiferroic, showing different kinds of magnetic ordering at low temperatures and partly also revealing significant magnetoelectric effects [12,13,14]. In certain regions of the rather complex magnetic phase diagrams of some lacunar spinels, even skyrmion-lattice states were found to emerge [17,18,19]. Skyrmions are whirl-like topological spin objects which are considered for applications in new types of data-storage devices [20,21,22,23]. In fact, the lacunar spinel GaV$_4$S$_8$ was the first bulk system where so-called Néel-type skyrmions were detected [17]. Interestingly, these magnetic objects in the lacunar spinels seem to be dressed with local electric polarization [13,18].

Ferroelectrics usually exhibit characteristic dipolar dynamics that can be detected by dielectric and/or optical experiments. Its investigation reveals crucial information, e.g., on the microscopic nature of the polar order, which, in conventional ferroelectrics, can arise from the displacement of ions (displacive ferroelectrics) or the ordering of permanent dipole moments (order-disorder ferroelectrics). Especially, in the latter case, where molecular or ionic rearrangements in multiwell potentials lead to relaxational processes, critical slowing down of the dipolar dynamics when approaching the ferroelectric transition at $T_{FE}$ is expected both above and below the transition [24]. Interestingly, for the JT-driven ferroelectric state of the lacunar spinels GaV$_4$S$_8$ and GeV$_4$S$_8$, such dipolar relaxation dynamics was recently detected, too [25,26]. Moreover, the peak in the temperature dependence of the low-frequency dielectric constant $\varepsilon'(T)$, commonly arising at ferroelectric transitions, was found to become successively suppressed for increasing measurement frequency [12,13,14], another typical feature of order-disorder ferroelectrics. In lacunar spinels, with the common formula $AB_4X_8$, the orbital ordering is accompanied by a distortion of the $B_4X_4$ cubane units, weakly linked molecule-like entities forming an fcc lattice. They behave like molecular magnets with well-defined orbital and spin degrees of freedom [27,28,29,30,31,32] and their orbital-order-induced distortion is believed to produce a dipolar moment, whose ordering causes the observed ferroelectricity [12,13]. Thus, in analogy to canonical order-disorder ferroelectrics, where the dipoles involved in the polar order already exist above $T_{FE}$, it can be assumed that the cubane units are already distorted above the JT transition This corresponds to a dynamic JT effect at $T > T_{JT}$.

Until now, GaV$_4$S$_8$ is the only lacunar spinel where the dipolar relaxation dynamics related to the ferroelectric state could be identified both below and above $T_{JT}$ [25]. Astonishingly, at the JT transition the corresponding relaxation time $\tau$ changes by more than four decades, becoming unusually short (of the order of 10$^{-13}$ s) at $T > T_{JT}$. This was ascribed to the first-order nature of the cooperative JT distortion, which prevents the divergence of $\tau(T)$, following a critical slowing down with $T_c < T_{JT}$ [25]. In GaV$_4$S$_8$, the distortion of the cubane units essentially involves an increase of the distance between one of the four V atoms and the opposite V$_3$-tetrahedron face. This is accompanied by a shrinking of the area of this V$_3$ face [27]. It is this stretching of the vanadium-clusters that leads to a dipolar moment and to ferroelectric polarization when long-range orbital order sets in. In marked contrast, in GaMo$_4$S$_8$ there is a *decrease* of this corner-face distance in the Mo$_4$ tetrahedra and the triangular face area grows [27]. As discussed in detail in Ref. [27], this difference of the two systems can be understood when considering that the highest molecular orbital in the V$_4$ clusters is occupied by one unpaired electron instead of one hole for the Mo$_4$ clusters. Via field-dependent polarization measurements, GaMo$_4$S$_8$ was very recently shown to become ferroelectric below its JT transition at 47 K and to exhibit switchable polarization [15]. We have further confirmed ferroelectricity in this materials by performing so-called positive-up-negative-down measurements [33]. Indeed JT-induced ferroelectricity in GaMo$_4$S$_8$ was theoretically predicted, also explaining the ferroelectricity in GaV$_4$S$_8$ [16]. Moreover, complex nanoscale ferroelectric-domain patterns were recently revealed in this compound by multiple scanning-probe microscopy [15]. GaMo$_4$S$_8$ is a type-I multiferroic, exhibiting a magnetic transition at about 20 K [34].

---

*Corresponding author:
peter.lunkenheimer@physik.uni-augsburg.de

In the present work, we provide a thorough characterization of GaMo$_4$S$_8$ both above and below $T_{JT}$ by broadband dielectric spectroscopy. We find complex and unusual dipolar relaxation dynamics, clearly distinct to that in GaV$_4$S$_8$. Above the JT transition, short-range relaxor ferroelectricity seems to compete with critical single-dipole fluctuations before the orbital ordering at $T_{JT}$ drives the system into the long-range ferroelectric order.

## II. EXPERIMENTAL DETAILS

GaMo$_4$S$_8$ single crystals were prepared by the flux method in a sealed molybdenum tube [35]. For the dielectric measurements, coplanar silver-paint contacts with a contact gap of 0.1 mm were applied on an as-grown (111) surface of the crystal. The resulting electrical field was oriented along the crystallographic [110] direction. For estimating the magnitudes of the real parts of the dielectric constant ($\varepsilon'$) and the conductivity ($\sigma'$), the penetration-depth of the field was assumed to be equal to the contact distance. Broadband measurements covering up to 10 frequency decades were performed by combining two techniques: Frequency-response analysis using a Novocontrol Alpha Analyzer (0.1 Hz - 3 MHz) and a coaxial reflectometric technique employing an Agilent E4991A impedance analyzer (1 MHz - 1.6 GHz) [36]. Data between 4.2 K and room temperature were taken in a $^4$He-bath cryostat (Cryovac).

## III. RESULTS AND DISCUSSION

Figure 1 shows the temperature dependence of $\varepsilon'$ and $\sigma'$ for a selection of frequencies. [Due to $\sigma' \propto \varepsilon'' \nu$, the $\sigma'(T)$ curves of Fig. 1(b) also qualitatively reflect the temperature dependence of the loss $\varepsilon''(T)$]. The occurrence of various peaks and steps in the temperature dependence indicates rather complex relaxation behavior arising from several, partly superimposed contributions. At $T_{JT} \approx 47$ K, a clear anomaly shows up in both quantities. $\varepsilon'(T)$ exhibits a strong increase below the ferroelectric transition and a peak for the lower frequencies. Its rather abrupt appearance when crossing the transition reflects the fact that the polar order is primarily promoted by the orbital order instead of growing dipolar correlations when approaching the transition. This anomaly in $\varepsilon'(T)$ becomes successively suppressed with increasing frequency and essentially is no longer observed above some 100 MHz. This is typical for order-disorder ferroelectrics and such a suppression was also found for GaV$_4$S$_8$ [13,37] and other lacunar spinels [12,14].

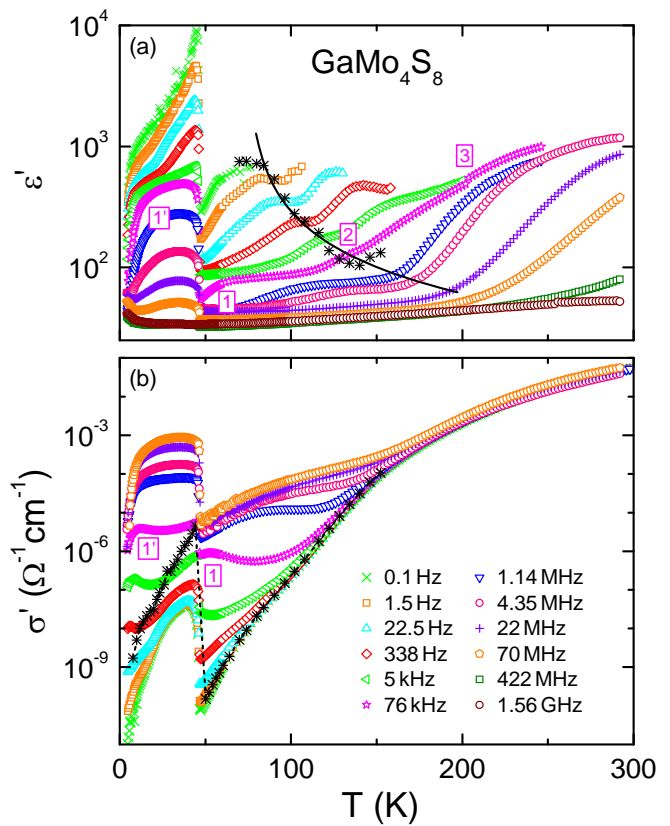

FIG. 1. Temperature dependence of $\varepsilon'$ (a) and $\sigma'$ (b) of GaMo$_4$S$_8$ at various frequencies. The stars in (a) and (b) represent $\varepsilon_s$ of the relaxor-like process and $\sigma_{dc}$, respectively, both obtained from the fits of the frequency-dependent data (Fig. 2). The line in (a) shows an approximate description of $\varepsilon_s(T)$ with a Curie-Weiss law, $\varepsilon_s \propto 1/(T - T_{CW})$ with $T_{CW} = 74$ K. The line in (b) is drawn to guide the eye. The numbers denote the observed relaxation features for the 76 kHz curve.

At $T > T_{JT}$ three separate relaxation processes can be identified, indicated by numbers 1-3 in Fig. 1(a) for the 76 kHz curve. They are revealed by the typical step-like increases in $\varepsilon'(T)$, shifting to higher temperatures with increasing frequency [38,39,40]. Similar to $GaV_4S_8$ [13] and $GaV_4Se_8$ [14], the process occurring at the highest temperatures, far above $T_{JT}$ (relaxation 3) is a so-called Maxwell-Wagner relaxation. It can be ascribed to electrode effects, commonly found for semiconducting materials due to depletion zones arising from the formation of Schottky diodes at the electrode-sample interfaces [41,42]. This is clearly confirmed by measurements of the same sample with different contact material, revealing a significant variation of $\varepsilon'(T)$ in this region [33]. In contrast, the other two relaxation processes detected above $T_{JT}$ (denoted 1 and 2 in Fig. 1) are of intrinsic nature. Relaxation 1, which is also signified by corresponding peaks in $\sigma'(T)$ [Fig. 1(b)], exhibits rather canonical characteristics with an upper plateau value (corresponding to the static dielectric constant $\varepsilon_s$) that moderately increases from about 50 to 90 with decreasing temperature. In marked contrast, relaxation 2 shows a very strong, Curie-Weiss like increase of $\varepsilon_s(T)$ upon cooling by more than one decade, reaching $\varepsilon_s$ values of the order of 1000 [cf. line in Fig. 1(a)]. The overall behavior of relaxation 2 reminds of that of a relaxor ferroelectric, where the dielectric behavior is usually explained in terms of short-range cluster-like ferroelectric order [43,44].

Below the JT transition, in addition to the rather sharp peak at $T_{JT}$, observed in the dielectric constant for low frequencies [Fig. 1(a)], at higher frequencies $\varepsilon'(T)$ exhibits a broad hump-like shape which becomes successively suppressed with increasing frequency. The decrease at the low-temperature flank of this hump seems to shift with frequency which indicates a relaxation process, denoted 1' in Fig. 1. Consistent with this finding, in $\sigma'(T)$ [and therefore also in $\varepsilon''(T)$] at the lowest temperatures and intermediate frequencies, a peak is observed, best visible for 5 and 76 kHz [Fig. 1(b)]. The corresponding loss peak is better resolved in the frequency-dependent data and will be discussed in more detail below.

Figure 2 shows the frequency dependence of $\varepsilon'$ and $\varepsilon''$ for various temperatures above (left column) and below the JT transition (right column). The $\varepsilon'$ spectra at $T > T_{JT}$ [Fig. 2(a)] exhibit up to three successive relaxation steps, best resolved and labeled for the 122 K curve. They correspond to relaxations 1-3 (from highest to lowest frequency), already identified in the $\varepsilon'(T)$ plot of Fig. 1(a). The corresponding loss peaks are well discernible for relaxation 1 as revealed by Figs. 2(c) and (e), the latter presenting a zoomed view of the peak region. However, for the other two relaxations, the $\varepsilon''$ peaks are not or only faintly visible. Obviously, they are strongly superimposed by the dc-conductivity contribution, $\varepsilon''_{dc} \propto \sigma_{dc}/\nu$, leading to an $1/\nu$ divergence at low frequencies, which also explains the absence of peaks for these relaxations in Fig. 1(b). Below $T_{JT}$ (right column of Fig. 2), in $\varepsilon'(\nu)$ two relaxation processes (termed 1' and 3') can be clearly identified. In the $\varepsilon''$ spectra, the dc contribution strongly superimposes the signature of process 3'.

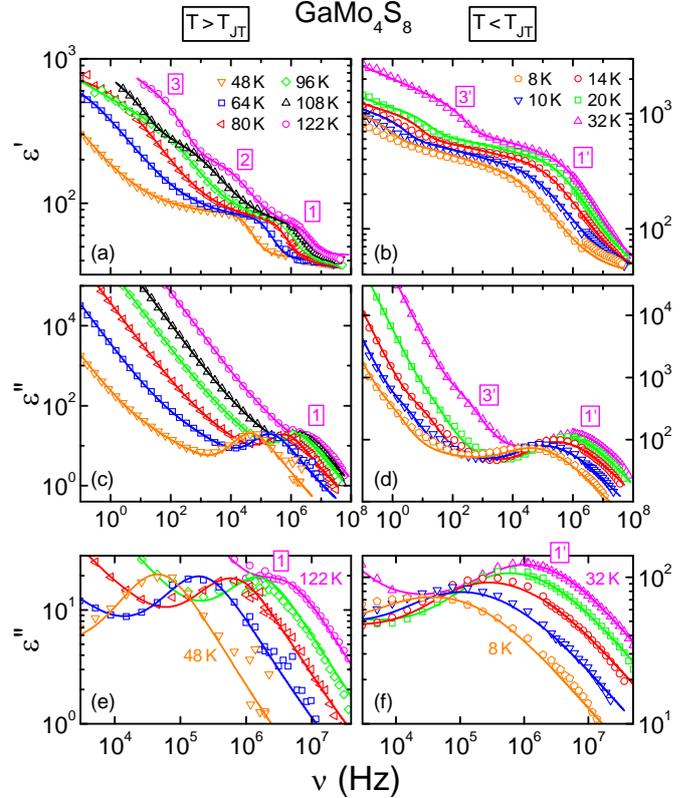

FIG. 2. Frequency dependence of $\varepsilon'$ (a,b) and $\varepsilon''$ (c-f) of $GaMo_4S_8$ at various temperatures. The left and right columns of the figure show the results above and below $T_{JT}$, respectively. Frames (e) and (f) represent zoomed views of the relaxation peaks. The solid lines are fits as described in the text. For 122 K, the numbers denote the different relaxation features observed above and below $T_{JT}$ (cf. Fig. 1).

To deconvolute the different relaxational contributions and to obtain quantitative information on the relaxation dynamics, simultaneous fits of the $\varepsilon'(\nu)$ and $\varepsilon''(\nu)$ spectra were performed, well reproducing the measured data (lines in Fig. 2). Here the electrode contributions were modeled by a distributed RC equivalent-circuit, assumed to be connected in series to the sample [45]. This was previously demonstrated to reasonably account for Maxwell-Wagner relaxations in numerous materials [42,45]. The two intrinsic processes, detected above $T_{JT}$, were fitted by the empirical Havriliak-Negami equation [46], commonly employed to describe relaxations in various classes of materials [26,38,40]. When adding a term for the dc conductivity, it reads as:

$$\varepsilon^* = \varepsilon' - i\varepsilon'' = \varepsilon_\infty + \sum_{j=1}^{2} \frac{\varepsilon_{s,j} - \varepsilon_{\infty,j}}{[1+(i\omega\tau_j)^{1-\alpha_j}]^{\beta_j}} - i\frac{\sigma_{dc}}{\varepsilon_0\omega} \qquad (1)$$

Here $j$ stands for the two intrinsic relaxations. $\varepsilon_\infty$ and $\varepsilon_0$ are the high-frequency limit of $\varepsilon'$ and the permittivity of vacuum, respectively. The width parameters $0 \leq \alpha < 1$ and $0 < \beta \leq 1$ in Eq. (1) generate a broadening of the relaxation step in $\varepsilon'(\nu)$ and of the peak in $\varepsilon''(\nu)$, compared to the Debye function. The latter corresponds to $\alpha = 0$ and $\beta = 1$ and can be derived when assuming that all dipoles move independently from each other and with an identical relaxation time. The frequently-observed broadening arises from a distribution of relaxation times due to disorder and/or interdipolar interactions [47,48], the latter being of special importance for crystalline materials [49]. Interestingly, in contrast to the other processes, relaxation 1 could be well fitted by a Debye function, i.e. with a single relaxation time, indicating that it is caused by non- or only weakly interacting dipoles. For relaxation 2, the fits result in a strongly temperature-dependent static dielectric constant [stars in Fig. 1(a)], in accord with its relaxor-ferroelectric characteristics discussed above. For $T < T_{JT}$, in addition to the Maxwell-Wagner relaxation (process 3'), a single intrinsic broadened relaxation (1') was already sufficient to fit the experimental data (right column of Fig. 2).

The fits also provide information on the intrinsic dc conductivity [stars in Fig. 1(b)], which shows a dramatic rise by more than four decades below $T_{JT}$. A similar jump of $\sigma_{dc}$ was also found for $GaV_4S_8$ [37]. The strong increase of $\sigma_{dc}$ below $T_{JT}$ may originate from charged domain walls which were observed by a recent force-microscopy study [15].

How can we interpret the intrinsic relaxation processes, detected in $GaMo_4S_8$? Figure 3(a) shows the temperature dependence of their mean relaxation times as obtained from the fits. Below $T_{JT}$, $\tau(T)$ of process 1' reveals thermally activated dynamics with a rather small energy barrier of $E = 3.4$ meV [Fig. 3(b)]. Interestingly, a low-temperature relaxation with small $E = 4.3$ meV was also observed in $GaV_4Se_8$ but not investigated up to the ferroelectric transition [14]. Just below $T_{JT}$, $\tau(T)$ of $GaMo_4S_8$ exhibits a small but significant increase with increasing temperature [Fig. 3(b) for $100/T < 3$]. A slowing down of relaxation dynamics when approaching the transition under heating is typical for order-disorder ferroelectrics [24,50,51] and was also found for $GaV_4S_8$ [25] and $GeV_4S_8$ [26]. Above $T_{JT}$, $\tau(T)$ of relaxation 1 is of similar order as the relaxation time of process 1' below $T_{JT}$. It exhibits critical behavior, $\tau \propto 1/(T-T_c)^\gamma$ [Fig. 3(c)], again as expected for order-disorder ferroelectrics [24,50], albeit with rather large critical exponent of $\gamma \approx 3$. Moreover, this relaxation is of Debye type as also commonly found for order-disorder ferroelectrics [24,50,52]. Therefore, relaxation 1 can be ascribed to the dipole dynamics expected for order-disorder ferroelectrics above $T_{FE}$. It arises from local fluctuations of single dipoles. Notably, in the present case of orbital-order driven ferroelectricity it also corresponds to orbital fluctuations and a dynamic JT effect above $T_{JT}$. The non-canonical jump-like discontinuity of $\tau(T)$ at $T_{JT}$ mirrors the first-order nature of the transition and the fact that, just as in $GaV_4S_8$, the ferroelectric polarization is primarily driven by the JT distortion. Below $T_{JT}$, relaxation 1' reflects coupled dipolar and orbital motions within the ferroelectric/orbital domains. Heterogeneities and domain oscillations in the ferroelectric phase, leading to a distribution of relaxation times, are plausible explanations of the non-Debye character of this process. While critical fluctuations govern its dynamics directly below $T_{JT}$, leading to a positive $\tau(T)$ gradient, at low temperatures obviously thermal activation becomes dominant.

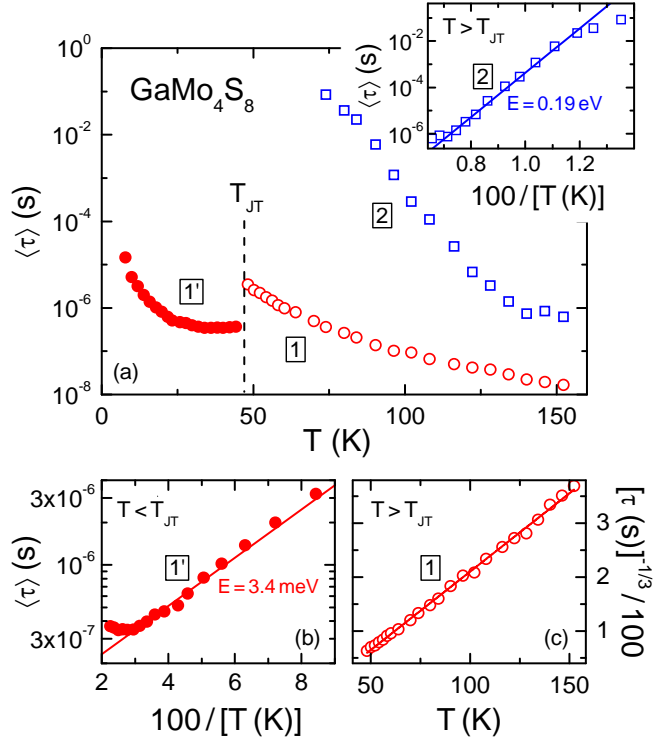

FIG. 3. (a) Temperature dependence of the average relaxation times of the intrinsic processes of $GaMo_4S_8$. The inset shows $\langle\tau\rangle$ of process 2 in Arrhenius representation. (b) Arrhenius representation of $\langle\tau\rangle$ of relaxation 1' below $T_{JT}$. (c) Critical behavior, $\langle\tau\rangle \propto 1/(T-T_c)^\gamma$, of process 1 ($T_c = 28$ K, $\gamma = 3$). All solid lines are linear fits.

In Ref. [25], similar microscopic mechanisms were assumed to interpret the dipolar relaxation dynamics in $GaV_4S_8$. However, there is one marked difference between the two systems: Above $T_{JT}$, for $GaV_4S_8$ the relaxational dynamics is exceptionally fast, with $\tau$ of the order of $10^{-13}$ s, detected by THz spectroscopy. In contrast, in $GaMo_4S_8$ these coupled dipolar and orbital fluctuations are about 5-7 decades slower and, thus, detectable by classical dielectric spectroscopy, just as for many other order-disorder ferroelectrics [24]. This difference in the relaxation times of the two lacunar spinels indicates that the assignment of single dipoles to the cubane units may be oversimplified and that the coupling of these units to the rest of the unit cell can be important. This is indeed manifested in the stronger JT distortion of $GaMo_4S_8$, which, moreover, involves a compression instead of a stretching of the tetrahedral cubane units, as described above [27].

What remains to be clarified is the nature of the relaxor-ferroelectric like behavior of $\varepsilon'(T)$ above $T_{JT}$ (relaxation 2). The relaxation time of this process behaves thermally activated in most of the investigated temperature range, with an energy barrier of 0.19 eV (inset of Fig. 3). In the whole temperature range, it is several decades slower than relaxation 1 [Fig. 3(a)]. Relaxor ferroelectricity is usually ascribed to short-range cluster-like ferroelectric order. Thus we propose the following scenario: Relaxation 2 signifies the dynamics of ferroelectrically correlated dipoles above $T_{JT}$, being slow and polydispersive (i.e., non-Debye) due to their strong interactions. While the majority of the dipoles in $GaMo_4S_8$ participates in this short-range polar order, existing already above $T_{JT}$, the much weaker relaxation 1 essentially mirrors single-dipole dynamics, corresponding to the orbital fluctuations preceding the JT transition. Without this transition, $GaMo_4S_8$ would be a relaxor ferroelectric but the JT transition finally drives the system into long-range ferroelectric order.

## IV. SUMMARY AND CONCLUSIONS

In summary, our thorough dielectric investigation of the multiferroic lacunar spinel $GaMo_4S_8$ has revealed astonishingly complex coupled dipolar and orbital relaxation dynamics with two intrinsic dynamic processes above and a single one below the JT transition. Our results suggest that, above $T_{JT}$, two types of dipolar fluctuations prevail: (i) Debye-like critical dynamics of only weakly interacting dipoles, as expected for order-disorder ferroelectrics, which here is closely coupled to the orbital fluctuations above $T_{JT}$. (ii) Much slower relaxations within cluster-like regions formed by dipoles with short-range ferroelectric correlations as commonly found for relaxor ferroelectrics. However, when long-range orbital order sets in below $T_{JT}$, finally long-range ferroelectricity is induced. Below this transition, dipolar dynamics within the ferroelectric/orbital domain structure is observed, again typical for order-disorder ferroelectricity. Remarkably, the coupled dipolar/orbital dynamics of $GaMo_4S_8$ strongly differs from that of $GaV_4S_8$, which above the JT transition exhibits a single, very fast Debye process in the sub-picosecond range.

## ACKNOWLEDGMENTS

We thank Alois Loidl for helpful discussions. This work was supported by the Deutsche Forschungsgemeinschaft through the Transregional Collaborative Research Center TRR 80.

# Supplemental Material

## Orbital-Order Driven Ferroelectricity and Dipolar Relaxation Dynamics in Multiferroic GaMo$_4$S$_8$

K. Geirhos[1], S. Krohns[1], H. Nakamura[2], T. Waki[2], Y. Tabata[2], I. Kézsmárki[1], and P. Lunkenheimer[1,*]

[1]*Experimental Physics V, Center for Electronic Correlations and Magnetism, University of Augsburg, 86159 Augsburg, Germany*
[2]*Department of Materials Science and Engineering, Kyoto University, Kyoto 606-8501, Japan*

*e-mail: Peter.Lunkenheimer@Physik.Uni-Augsburg.de

## 1. Positive-up-negative-down measurements

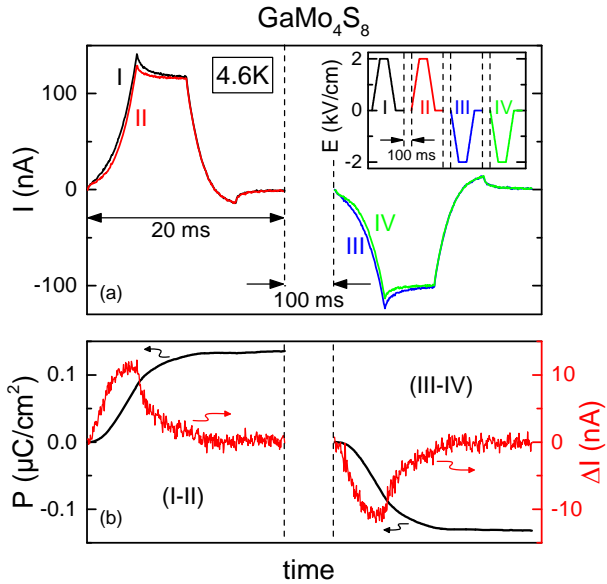

FIG. S1. (a) Time-dependent current obtained from PUND measurements performed at 4.6 K with a waiting time of 100 ms and a pulse width of 20 ms as indicated. For a better comparison, pulses I and II and pulses III and IV are shown with superimposed time scales. Inset: Electric-field excitation signal. (b) Red lines (right ordinate): Difference of pulses I and II (left part of the figure) and pulses II and IV (right). Black lines (left ordinate): Resulting polarization $P(t)$.

To further corroborate the recently reported ferroelectricity in GaMo$_4$S$_8$ below the Jahn-Teller (JT) transition [15], here we present so-called positive-up-negative-down (PUND) measurements (Fig. S1). In this type of experiment, the sample is subjected to a sequence of trapezoid field pulses (inset of Fig. S1) and the resulting time-dependent current $I(t)$ is monitored. As revealed in Fig. S1(a), for GaMo$_4$S$_8$ the main effect detected in $I(t)$ is an approximate proportionality to the applied voltage pulses, which simply arises from charge transport within the semiconducting sample [cf. dc-resistivity $\sigma_{dc}(T)$ shown by the stars in Fig. 1(b)]. The flanks of the measured $I(t)$ trapezoids show a slight curvature and some over/undershoots, which can be explained in terms of the trivial charging/discharging dynamics of the sample capacitor. However, a close inspection of Fig. S1(a) reveals some significant differences of $I(t)$ for pulses I and II and for pulses III and IV. The difference curves are shown in Fig. S1(b) (red lines). Obviously, at the increasing flanks of the first and third pulses, an enhancement of $I(t)$ arises. Such difference of the response to two succeeding pulses is typical for ferroelectric polarization: During the first and third pulse, the field leads to a switching of the macroscopic polarization, generating a reorientation of the dipolar moments within the ferroelectric domains and, thus, an excess contribution to $I(t)$. However, at the second and forth pulses, having the same polarity as the preceding ones, the polarization was already switched by the previous pulse and no further dipolar reorientation (and, thus, no excess $I$) is expected. From an integration of the detected excess currents, the polarization can be calculated [black lines in Fig. S1(b)]. The obtained values are of the same order of magnitude as $P$ deduced from pyrocurrent measurements reported in Ref. [15].

## 2. Dependence of $\varepsilon'$ on electrode material.

Figure S2 shows the temperature dependence of the dielectric constant $\varepsilon'$ at various frequencies, measured with two types of electrode material, namely sputtered gold (lines) and silver paint (symbols). The relaxation steps observed at the highest temperatures obviously significantly change for the different types of electrodes. This shows that in this region the dielectric response is completely of non-intrinsic origin and the observed relaxation is of Maxwell-Wagner type [41,42,45]. In contrast, the two relaxation features observed at lower temperatures (but still above $T_{JT}$) reasonably agree for both measurements, confirming their intrinsic nature. Below the JT transition, the situation is more complex. The $\varepsilon'(T)$ curves disagree at low frequencies but well agree at the highest frequency. This indicates that an equivalent-circuit description, involving a shortening of the contact capacitor at high frequencies, should be able to reveal the intrinsic properties of the sample material.

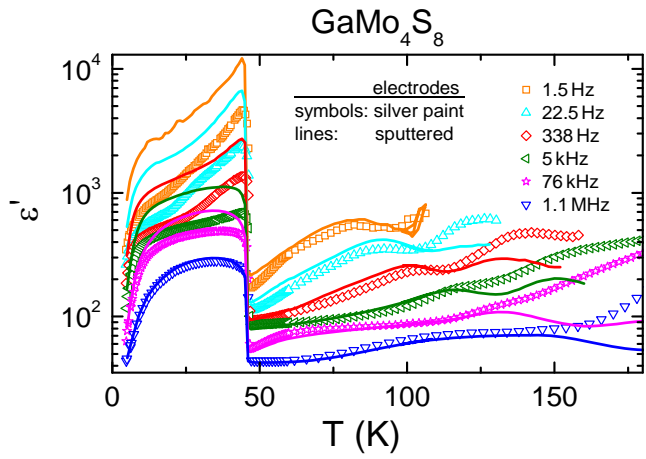

FIG. S2. Temperature dependence of $\varepsilon'$ of GaMo$_4$S$_8$ measured at various frequencies with two different types of contact.